\documentclass[english,aps,prl,superscriptaddress,twocolumn,showpacs,footinbib]{revtex4-1}
\usepackage{graphicx}  
\usepackage{dcolumn}   
\usepackage{bm}        
\usepackage{amssymb}   
\usepackage{babel}
\usepackage{float}
\usepackage{verbatim}
\usepackage{amsmath}
\usepackage{setspace}
\usepackage{epstopdf}
\graphicspath{{figures/}}
\usepackage{siunitx}
\usepackage{hyperref}
\usepackage{mathtools}

\DeclareMathOperator{\Tr}{Tr}
\DeclareMathOperator{\PS}{PS}

\begin{document}
\widetext

\title{Hamiltonian engineering of general two-body spin-1/2 interactions}

\author{K. I. O. Ben 'Attar} \thanks{These authors contributed equally to this work, kevin.benattar@outlook.com} 
\affiliation{Dept. of Applied Physics, Rachel and Selim School of Engineering, Hebrew University, Jerusalem 9190401, Israel}

\author{D. Farfurnik} \thanks{These authors contributed equally to this work, Corresponding author, dimka\_f13@yahoo.com}
\affiliation{Dept. of Applied Physics, Rachel and Selim School of Engineering, Hebrew University, Jerusalem 9190401, Israel}
\affiliation{Racah Institute of Physics, The Hebrew University of Jerusalem, Jerusalem 9190401, Israel}

\author{N. Bar-Gill}
\affiliation{Dept. of Applied Physics, Rachel and Selim School of Engineering, Hebrew University, Jerusalem 9190401, Israel}
\affiliation{Racah Institute of Physics, The Hebrew University of Jerusalem, Jerusalem 9190401, Israel}

%

\date{\today}

\begin{abstract}
Spin Hamiltonian engineering in solid-state systems plays a key role in a variety of applications ranging from quantum information processing and quantum simulations to novel studies of many-body physics. By analyzing the irreducible form of a general two-body spin-1/2 Hamiltonian, we identify all interchangeable interaction terms using rotation pulses. 
Based on this identification, we derive novel pulse sequences, defined by an icosahedral symmetry group, providing the most general achievable manipulation of interaction terms. We demonstrate that, compared to conventional Clifford rotations, these sequences offer advantages for creating Zeeman terms essential for magnetic sensing, and could be utilized to generate new interaction forms. The exact series of pulses required to generate desired interaction terms can be determined from a linear programming algorithm. For realizing the sequences, we propose two experimental approaches, involving pulse product decomposition, and off-resonant driving. Resulting engineered Hamiltonians could contribute to the understanding of many-body physics,  and result in the creation of novel quantum simulators and the generation of highly-entangled states, thereby opening avenues in quantum sensing and information processing.
\end{abstract}
\maketitle

\section{I. Introduction}
\paragraph{}
For many decades, rotation pulse sequences have been utilized in nuclear magnetic resonance (NMR) for manipulating spin states through Hamiltonian engineering \cite{Waugh1968,Mansfield1971}. In recent years, solid-state spin systems such as defects in diamond, silicon, and silicon carbide, as well as atomic systems (including Rydberg atoms and tweezer-trap arrays), have emerged as useful platforms for quantum technologies, thereby reviving the neccesity in efficient spin control schemes. In particular, Nitrogen-Vacancy (NV) centers in diamond, which offer optical initialization and readout capabilities, and can be treated to some extent as spin-1/2 qubits, are widely used for sensing \cite{Taylor2008,maze2008,Balasubramanian2008,Dolde2011,Barry2016,Chatzidrosos2017,Farfurnik2018} and quantum information processing \cite{Jelezko2006,Bernien2013,Tsukanov2013,Hensen2015}. Manipulating the dipolar interactions within an ensemble of such spins could pave the way towards novel studies of many-body dynamics \cite{Choi2017,ChoiS2017,Farfurnik2018dip}, the creation of quantum simulators and advanced quantum sensors \cite{Okeeffe2019}, and generation of non-classical spin states \cite{Cappellaro2009, Farfurnik2018dip}. Recent studies of such Hamiltonian engineering, analyzing the effects of control pulses from the Clifford rotation group, resulted in a novel scheme of generating certain types of Hamiltonians \cite{Okeeffe2019}. 

\paragraph{}

Here, we use group theory to go beyond previous work, introducing a completely general platform for interaction manipulation, namely pulse sequences defined by an icosahedral symmetry. While describing the limitations of achievable control with standard Clifford-based rotations, we emphasize that icosahedral pulses allow the most general achievable engineering of Hamiltonian interaction terms. In the practical case of quantum sensing in interacting ensembles, this approach is shown to break the performance bound of the standard Clifford-based rotation schemes. By utilizing a linear programming algorithm, we derive the proper sequences for transforming the natural NV-NV dipolar interaction Hamiltonian to several target Hamiltonians, some of which could not be previously obtained. Our proposal of two practical experimental approaches for realizing these schemes, and supporting simulations of the resulting spin dynamics, emphasize the applicability of the icosahedral sequences in a variety of quantum systems consisting of spin ensembles.

\section{II. Irreducible Interaction Representation}
\paragraph{}  
We begin by introducing the general interaction Hamiltonian of $N$ spin-1/2 particles, which contains up to two-body interaction terms,
\begin{equation}
H = \sum_{a,i} \vec{n}_i^{(a)} \cdot \vec{\sigma}_i^{(a)}  + \sum_{a<b,i,j}T^{(ab)}_{ij}\sigma^{(a)}_i \sigma^{(b)}_j,
\label{general}
\end{equation}
where $\sigma$ represents the Pauli spin operators, $a,b\in\{1\dots N\}$ the spin indices in the ensemble, $T$ ($\vec{n}$) is the matrix (vector) of coefficients for the two-body interaction (one-spin) terms, and $i,j\in \{x,y,z\}$. The coefficient matrix can be written in the irreducible form 
\small
\begin{align}
\label{eq:irreduc}
 & T_{ij}^{(ab)}\sigma^{(a)}_i \sigma^{(b)}_j =  
\alpha^{(ab)} \vec{\sigma}^{(a)} \cdot \vec{\sigma}^{(b)} \nonumber\\ &+\vec{\beta}^{(ab)}\cdot   
\underbrace{\begin{bmatrix}
	\sigma^{(a)}_y \sigma^{(b)}_z - \sigma^{(a)}_z \sigma^{(b)}_y \\
	\sigma^{(a)}_z \sigma^{(b)}_x - \sigma^{(a)}_x \sigma^{(b)}_z \\
	\sigma^{(a)}_x \sigma^{(b)}_y - \sigma^{(a)}_y \sigma^{(b)}_x
\end{bmatrix}}_{\vec{\lambda}^{(ab)}_1}
+\vec{\gamma}^{(ab)}\cdot
\underbrace{\begin{bmatrix}
	\sigma^{(a)}_x \sigma^{(b)}_x - \sigma^{(a)}_z \sigma^{(b)}_z \\
	\sigma^{(a)}_y \sigma^{(b)}_y - \sigma^{(a)}_z \sigma^{(b)}_z \\
	\sigma^{(a)}_x \sigma^{(b)}_y + \sigma^{(a)}_y \sigma^{(b)}_x \\
	\sigma^{(a)}_x \sigma^{(b)}_z + \sigma^{(a)}_z \sigma^{(b)}_x \\
	\sigma^{(a)}_y \sigma^{(b)}_z + \sigma^{(a)}_z \sigma^{(b)}_y
\end{bmatrix}}_{\vec{\lambda}^{(ab)}_2},
\end{align}
\normalsize
with $\alpha^{(ab)} = \frac{\Tr{T}^{(ab)}}{3}$, $\beta^{(ab)}_k=\frac{1}{2}\epsilon_{ijk}T^{(ab)}_{ij}$ and $\vec{\gamma}^{(ab)}$ is extracted from the traceless symmetric matrix $\left(\frac{ T+T^{\dagger}}{2} -  \frac{\Tr{(T)}}{3} \right)=\begin{bmatrix}
\gamma_1 & \gamma_3 & \gamma_4 \\
\gamma_3 & \gamma_2 & \gamma_5 \\
\gamma_4 & \gamma_5 & -\gamma_1-\gamma_2 
\end{bmatrix}$ by $\vec{\gamma} = \begin{bmatrix}
 \gamma_1\\
 \gamma_2\\
 \gamma_3\\
 \gamma_4\\
 \gamma_5
\end{bmatrix}$. In the first order of the Magnus expansion \cite{Magnus1954}, applying a series of pulses $\{P_k\}$ at times $\{t_k\}$ can transform the initial Hamiltonian \eqref{general} to the engineered Hamiltonian 
\begin{align}
\label{eng}
& \bar{H} = \sum_{a,i}\PS^{(1)}(\vec{n}_i^{(a)})\cdot   
\vec{\sigma}^{(a)}_i+\sum_{a<b,i<j} \left[
\alpha^{(ab)} \vec{\sigma}^{(a)} \cdot \vec{\sigma}^{(b)}\right.\nonumber\\ & +\left.\PS^{(1)}(\vec{\beta}^{(ab)})\cdot   
\vec{\lambda}^{(ab)}_1
+\PS^{(2)}(\vec{\gamma}^{(ab)})\cdot
\vec{\lambda}^{(ab)}_2\right], 
\end{align}
where $\PS^{(1)}$ and $\PS^{(2)}$ are $3 \times 3$ and $5 \times 5$ matrices uniquely representing the effects of the pulse sequence on the interaction terms (Appendix A). These matrices correspond to convex combinations of the three dimensional matrix representing rotation around the axis $\hat{n}$ by a general angle $0<\theta<2\pi$, $R_{\hat{n},\theta}^{3\times 3}$, and of a $5 \times 5$ reduced form of $R_{\hat{n},\theta}^{3\times 3}\otimes R_{\hat{n},\theta}^{3\times 3}$ respectively (after removing the redundancy in the traceless symmetric matrix, see Appendix A). Here, we will focus on specific rotation angles $\theta$ and axes $\hat{n}$ characterizing two symmetry groups, the Clifford group and the regular icosahedral group, demonstrating that the latter provides the most general interaction manipulation capabilities. Note that rotation pulses cannot modify the isotropic interaction part $\vec{\sigma}^{(a)} \cdot \vec{\sigma}^{(b)}$.
\section{III. Hamiltonian Engineering Concepts}
\subsection{A. Clifford and Icosahedral Rotation Sequences}
\paragraph{}
Forming the Clifford rotation group, one basic family of spin control sequences consists of a series of $(\frac{\pi}{2})$-rotation pulses $\{P_k\}$ applied around the three principle axes of the Bloch-sphere at times $\{ t_k\}$ \cite{Okeeffe2019}. For adjacent pulses with zero spacings ($t_j-t_{j-1}=0$ for a given $j$), such a structure considers rotations by any integer multiplying angles of  $(\frac{\pi}{2}$).  
\begin{figure}[!t]	
	\includegraphics[width=1\columnwidth]{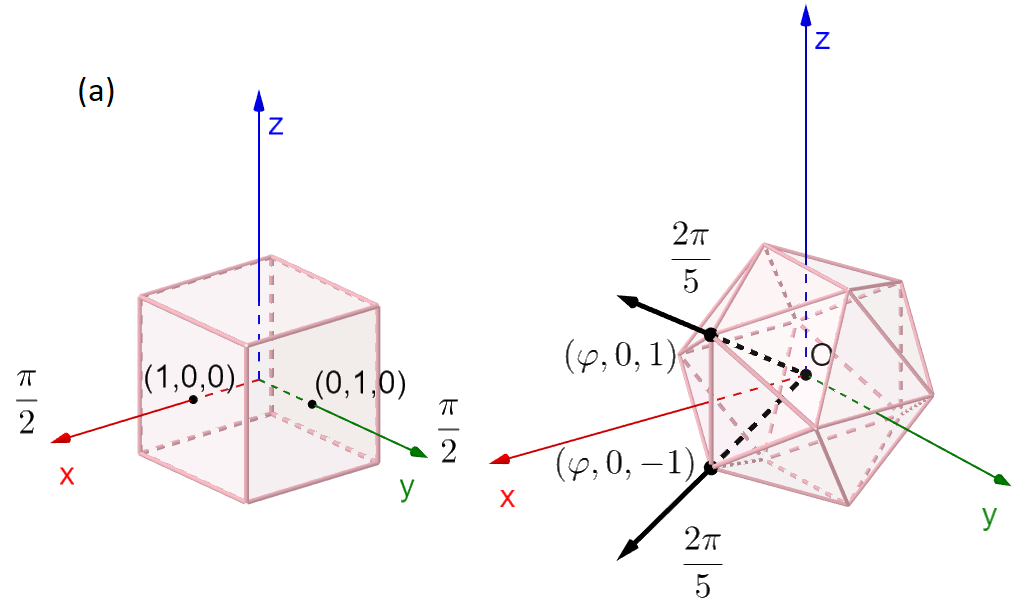}
	\caption{(Color online) Schematic of the angles and axes of pulses derived from the (a) Clifford and (b) icosahedral symmetry groups, with $\varphi=\frac{1+\sqrt{5}}{2}$ the golden ratio.}
	\label{fig:schemes}
\end{figure}
The novel utilization of such sequences, which were obtained by linear programming, was recently proposed for the engineering of the one-spin Hamiltonian proportional to $\sum_a \sigma^{(a)}_{z}$ \cite{Okeeffe2019}. We extend this work by studying the general Hamiltonian engineering capabilities of the Clifford group, as well as a higher dimensional icosahedral symmetry group. 
\paragraph{} 
In considering the structure of matrices forming the Clifford rotation group \cite{Cromwell1997}, although the general structure of $\PS^{(1)}$ can contain nine nonzero elements, the form of $\PS^{(2)}$ must consist of two independent ($2 \times 2$ and $3 \times 3$) blocks, with all other elements equal to zero. As a result, the capabilities of interchanging different interaction terms are limited. While the isotropic part $\vec{\sigma}^{(a)} \cdot \vec{\sigma}^{(b)}$ cannot be manipulated at all, different interaction terms inside the vector $\vec{\lambda}^{(ab)}_1$ can be interchanged only among themselves, similarly to the first two components of $\vec{\lambda}^{(ab)}_2$ and the last three components of $\vec{\lambda}^{(ab)}_2$. For example, the initial interaction term $\sigma_x^{(a)}\sigma_x^{(b)}-\sigma_z^{(a)}\sigma_z^{(b)}$   cannot be transformed into the term $\sigma_x^{(a)}\sigma_y^{(b)}+\sigma_y^{(a)}\sigma_x^{(b)}$ by applying sequences consisting of solely $(\frac{\pi}{2})$-pulses.
\paragraph{}
The structure of matrices forming the regular icosahedral symmetry group \cite{Cromwell1997}, however, provides a different picture: the application of a series of $\left(\frac{2\pi}{5}\right)$-pulses, at angles determined by this symmetry group, imposes no limitations upon the elements of the matrix $\PS^{(2)}$. As a result, such icosahedral pulses enable interchanging different interaction terms inside the vector $\vec{\lambda}^{(ab)}_2$, thereby providing the most general manipluation capabilities of two spin-interaction terms. Utilizing icosahedral pulses could transfer an interaction term of the form $\sigma^{(a)}_x \sigma^{(b)}_x-\sigma_z^{(a)} \sigma_z^{(b)}$, for example, to the final term $\sigma^{(a)}_x \sigma^{(b)}_z+\sigma_z^{(a)} \sigma_x^{(b)}$, which could not be generated solely by Clifford rotations. 
\subsection{B. Extracting a Pulse Sequence for a Target Hamiltonian}
\paragraph{}
Extracting the right pulse sequence that transforms an initial Hamiltonian to a desired one involves utilizing linear programming \cite{Okeeffe2019}. Given a matrix $A$, a vector $\vec{b}$ and lower and upper bounds $\vec{l}_b, \vec{u}_b$, such an algorithm is designed to find the vector $\vec{x}$ satisfying
 \begin{equation}
 \label{prog}
 A\vec{x}=\vec{b} 
 \end{equation}
 between the limits $\vec{l}_b<\vec{x}<\vec{u}_b$, while minimizing the $l_1$ norm of the solution.  The problem of identifying the right sequence for Hamiltonian engineering is mapped onto this algorithm in the following way \cite{Okeeffe2019}: considering all $\{R_k\}$ matrices forming the chosen rotation symmetry group (Clifford or icosahedral in our case), and the vectors $\vec{v}_{in} = [\vec{\beta}_{in} \quad \vec{\gamma}_{in}]^T$ (or $[\vec{n}_{in} \quad \vec{\gamma}_{in}]^T$, see  below) and $\vec{b} = [\vec{\beta}_f  \quad\vec{\gamma}_f]^T$ (or $[\vec{n_f} \quad \vec{\gamma}_f]^T$)  representing the initial and final interaction terms [see Eq. \eqref{eq:irreduc}], we denote $A=(\vec{v}_i \otimes 1)R$, where $R$ is a matrix whose different columns represent the effects of different rotations from the symmetry group (Appendix A). This way, for a sequence with a total cycle time $T$, the $k$'th element of the solution vector to \eqref{prog}, $x_k=\frac{\tau_k}{T}$, will represent the duration within the sequence $\tau_k$ associated with the rotation $R_k$ (for excluded rotations, $x_k=0$ ). As the actual applied control pulses $\{P_k\}$ at times $\{t_k\}$ satisfy the relation  $R_k=\prod_{i=1}^{k} P_i$ (Appendix A), inverting this relation will identify the desired pulse sequence. Additional symmetrization was applied to cancel higher (even) orders of average Hamiltonian terms. Moreover, initial and final pulses were introduced to satisfy the cyclic condition of the average Hamiltonian theory for a sequence with $n$ pulses,  $\bar{H}_{1}=\bar{H}_{n}$ \cite{Magnus1954,Waugh1968,Mansfield1971}, and pulses with nontrivial angles and axes were decomposed (as composite pulses) to products of $\left(\frac{\pi}{2}\right)$ and $\left(\frac{2\pi}{5}\right)$-pulses.
 \begin{figure*}[!t]	
 	\includegraphics[width=2\columnwidth]{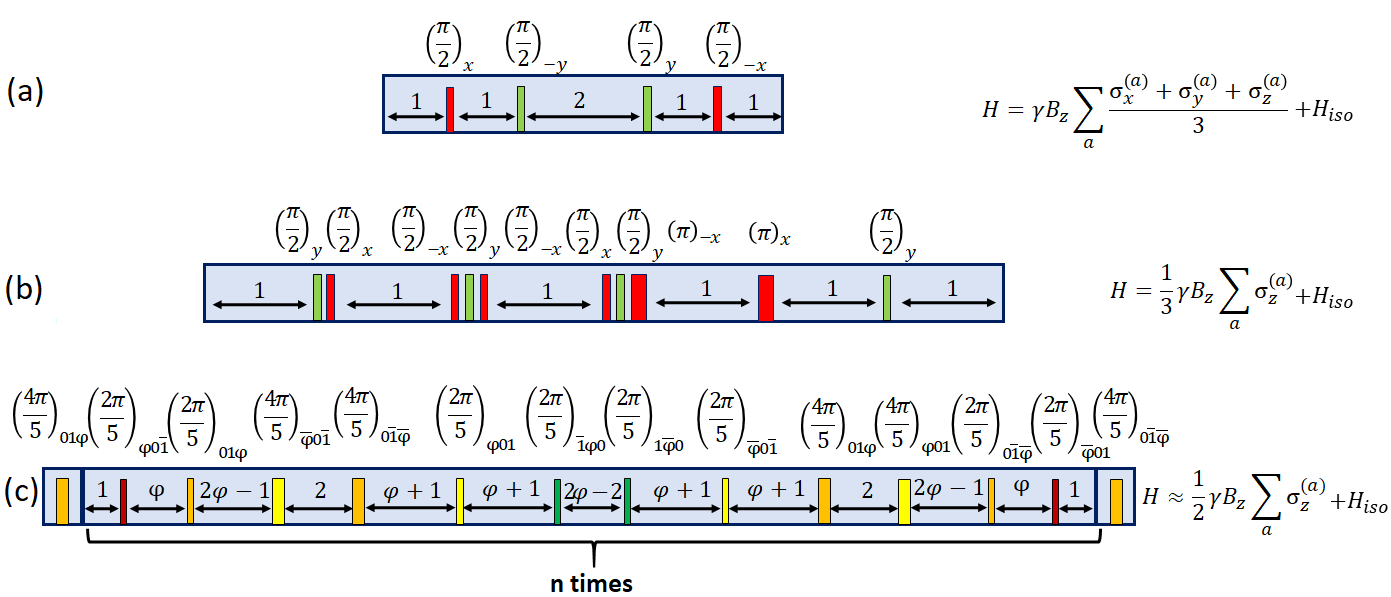}
 	\caption{(Color online) 	Pulse sequences generating various target Hamiltonians (right column) from the initial NV-NV interaction Hamiltonian \eqref{eq:NVNV}: (a) Averaged out dipolar interactions by utilizing the WAHUHA sequence. (b-c) Zeeman Hamiltonian essential for magnetic sensing, utilizing pulses taken from the (b) Clifford \cite{Okeeffe2019} and (c) icosahedral symmetry groups. The icosahedral sequence results in a stronger Zeeman term. Rotation angles and axes are presented above each schematic, time spacings between pulses are given in arbitrary units.}
 	\label{fig:sequences1}
 \end{figure*}
 \section{IV. Implementation on Ensembles of NV Centers in Diamond}
\paragraph{}
One of the most promising platforms for studying many-body dynamics in the solid-state is the optically addressable spin ensemble of NV centers. We now discuss, within the framework of the average Hamiltonian theory, several useful target Hamiltonians that could be engineered by pulse sequences affecting the initial NV-NV interaction Hamiltonian (essentially relevant for general dipole coupling). Considering a two-level manifold (e.g. $m_s=0,1$) of the NV ground states with a particular crystallographic orientation, in the rotating frame with respect to the bare Hamiltonian, and under the rotating wave approximation, such an interaction Hamiltonian is given by \cite{Choi2017,Kucsko2018,Farfurnik2018dip}
\begin{equation}
\label{eq:NVNV}
H_{NV}= \gamma B_z\sum_a{\sigma^{(a)}_{z}}+\sum_{a<b}\omega^{(ab)}\left(\vec{\sigma}^{(a)}\cdot\vec{\sigma}^{(b)}-2\sigma_z^{(a)}\sigma_z^{(b)}\right) ,
\end{equation}
where the dipolar interaction strengths $\omega^{(ab)}$ between spins $a$ and $b$ depend on their relative positions, $\gamma$ is the electron gyromagnetic ratio, and $B_z$ corresponds to an external magnetic field. The second term of the right-hand side of \eqref{eq:NVNV} represents dipolar interactions, while the first term, often referred to as the ``Zeeman term", is essential for magnetic sensing. 
\subsection{A. Reproduction of Known Sequences}
First, we verify the validity of the linear programming procedures by reproducing the well-known WAHUHA sequence \cite{Waugh1968} [Fig. \ref{fig:sequences1} (a)], which was designed in NMR to decouple spin-1/2 dipolar interaction, while flipping the Zeeman term along the three axes of the Bloch-sphere. Due to the inequivalency between the two-level manifold of NV centers and spin-1/2 systems, the dipolar terms are not decoupled completely, and the isotropic part $H_{iso}=\sum_{a<b}\omega^{(ab)}\frac{\vec{\sigma}^{(a)}\cdot \vec{\sigma}^{(b)}}{3}$ remains \cite{Kucsko2018,Farfurnik2018dip} [Fig. \ref{fig:sequences1} (a)]. Modifying the timing of pulses in the WAHUHA sequence could result in the generation of novel interaction Hamiltonians, paving the way towards the creation of non-classical states such as spin-squeezed states \cite{Cappellaro2009,Farfurnik2018dip}. 
\paragraph{}
Subsequently, we utilize the linear programming algorithm for extracting the sequences required for generating several useful Hamiltonians (Figs. \ref{fig:sequences1}, \ref{fig:sequences2}). By considering Clifford rotations only, our code reproduces the HoRD-qubit-5 pulse sequence obtained by O'Keeffe et al. \cite{Okeeffe2019} [Fig. \ref{fig:sequences1} (b)], for decoupling dipolar interactions while keeping the one-spin Zeeman Hamiltonian, which is essential for magnetic sensing. 
\subsection{B. Novel Hamiltonian Engineering Utilizing Icosahedral Sequences}
\begin{figure}[!t]	
	\includegraphics[width=1\columnwidth]{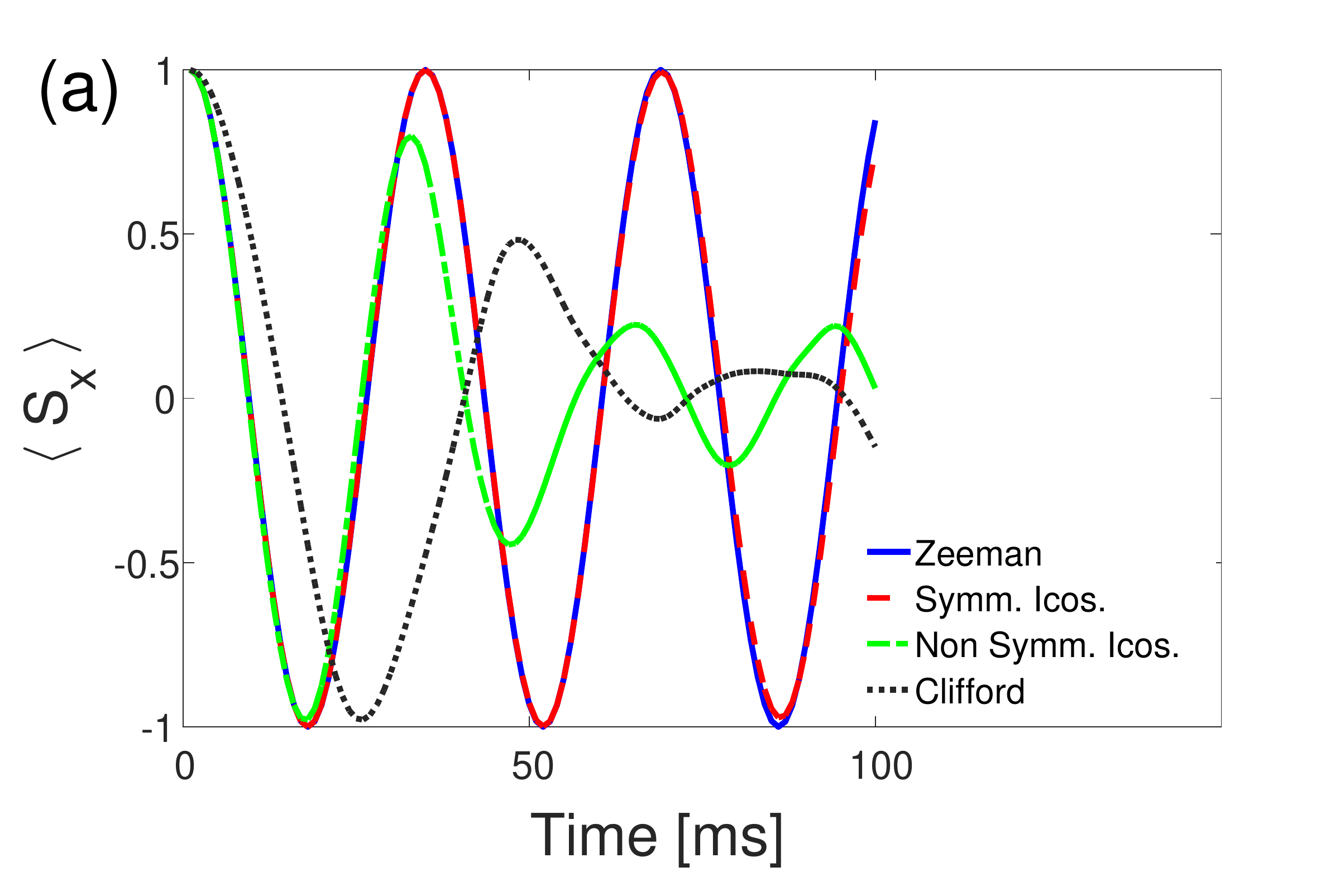}
	\caption{(Color online) Simulated fidelity of an NV ensemble spin state initialized along the x axis as a function of time, under the evolution of an ideal Zeeman Hamiltonian with $B_z= 2$ nT, and target Zeeman Hamiltonians generated by 100 repetitions of sequences presented in Fig. \ref{fig:sequences1} (b) and (c): Clifford rotations, symmetrized and non symmetrized icosahedral sequences. The 1/2 Zeeman coefficients generated by icosahedral pulses could offer enhanced sensitivities over Clifford rotations, which result in a Zeeman coefficient of 1/3.}
	\label{fig:dynamics}
\end{figure}

\begin{figure*}[t!]	
 	\includegraphics[width=2\columnwidth]{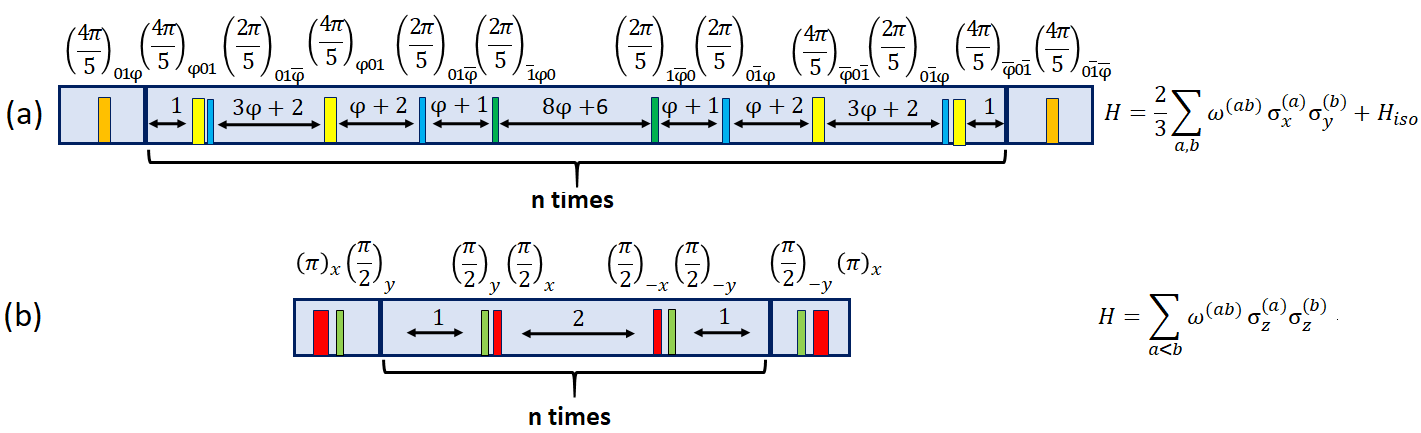}
 	\caption{(Color online) 	Pulse sequences generating various target Hamiltonians (right column) from the initial NV-NV interaction Hamiltonian \eqref{eq:NVNV} with $B_z=0$: (a) $\sigma_x \sigma_y$ Hamiltonian, which could not be generated solely by Clifford rotations. (b) $\sigma_z$-product Hamiltonian excluding any additional isotropic terms, under which spin dynamics can be expressed analytically.(b)  Rotation angles and axes are presented above each schematic, time spacings between pulses are given in arbitrary units.}
 	\label{fig:sequences2}
 \end{figure*}

\paragraph{}
By extending spin control to pulses oriented along the angles of an icosahedron, we reveal a novel 14-pulse sequence consisting of $(\frac{2\pi}{5})$ and $(\frac{4\pi}{5})$-pulses, which is expected to generate a similar Zeeman term with strength greater than HoRD-qubit-5 ($\sim \frac{1}{2}$ versus $\frac{1}{3}$, as emphasized by simulations in Fig. \ref{fig:dynamics}). 
We now detail this derivation: starting from the initial NV Hamiltonian \eqref{eq:NVNV}, a two-body interaction term with strength $\omega$ can be written according to Eq. \eqref{eq:irreduc} in the irreducible form $\vec{n}_{in}=\gamma B_z[0,0,1]$, $\alpha_{in}=\frac{1}{3}\omega$, $\vec{\beta}_{in}=\vec{0}$, $\vec{\gamma}_{in}=\frac{2}{3}\omega[1,1,0,0,0]$. Similarly, the target Zeeman interaction corresponds to $\vec{n}_{f}=\gamma B_z[0,0,1]$, $\alpha_{f}=\frac{1}{3}\omega$, $\vec{\beta}_{f}=\vec{0}$, $\vec{\gamma}_{f}=\vec{0}$. Since  $\vec{\beta}=\vec{0}$ and the isotropic term is not affected by pulse sequences, the linear programming is solved for the remaining nonzero terms $\vec{v}_{in}=[\vec{n}_{in} \quad \vec{\gamma}_{in}]^T$, $\vec{v}_{f}=[\vec{n}_{f} \quad \vec{\gamma}_{f}]^T$. The resulting seven nonzero components of $\vec{x}$, $[1,\varphi,2\varphi-1,2,\varphi+1,\varphi+1,2\varphi-2]$, associated with  the rotation matrices with angles and axes presented in Fig. \ref{fig:sequences1} (c), are then symmetrized to from a 12-pulse sequence. Finally, applying additional initial and final $\left(\frac{4\pi}{5}\right)$-pulses around the angles $[0,1,\varphi]$, $[0,-1,-\varphi]$ provide the cyclic condition $\bar{H}_{1}=\bar{H}_{15}$ essential for Hamiltonian engineering \cite{Magnus1954,Waugh1968,Okeeffe2019}. The average Hamiltonian resulting from applying this sequence, $\bar{H}=\frac{\sum_{k=1}^{15} \tau_k \bar{H}_k}{T}$, incorporates a Zeeman coefficient of $\frac{1}{2}$ [Fig. \ref{fig:sequences1} (c)], which is greater than the $\frac{1}{3}$ coefficient produced by the HoRD-qubit-5 sequence. In Fig. \ref{fig:dynamics} we compare between simulated dynamics under the pure Zeeman Hamiltonian with the average Hamiltonians obtained using the HoRD sequence, the icosahedral sequence and its symmetrized variant. The clear advantage of the latter is evident. As a result, utilizing such icosahedral pulses could potentially result in enhanced magnetic sensitivities.
\paragraph{}
 The advantages of icosahedral pulse sequences can be further emphasized by applying a novel 10-pulse sequence [Fig. \ref{fig:sequences2}(a)] on the NV ensemble with no external field ($B_z=0$), resulting in the interaction term $\sigma_x^{(a)} \sigma_y^{(b)}$, which cannot be generated solely by Clifford rotations. Another useful interaction term, generatable at $B_z=0$ by either Clifford rotations or icosahedral pulses, is proportional to $\sigma_z^{(a)} \sigma_z^{(b)}$ [Fig. \ref{fig:sequences2}(b)], excluding additional isotropic terms. For a product state initialized along the x-y plane of the Bloch-sphere, general spin dynamics under this Hamiltonian can be written as a closed analytic expression \cite{Farfurnik2018dip}, providing deep physical understanding of the system, as well as simulation capabilities of large ensembles.
 \subsection{C. Experimental Realization Schemes}
 \paragraph{}
 The conventional experimental realization of NV spin control utilizes in-phase-and-quadrature (IQ) modulated microwave (MW) signal, generating on-resonant pulses along the $x-y$ plane of the Bloch-sphere. However, in contrast to conventional pulse sequences consisting of pulses solely along the $x-y$ plane, implementing the icosahedral sequences requires spin control along the $x-z$ and $y-z$ planes. One possible experimental realization of such pulses involves their decomposition into products of three rotations in the $x-y$ plane: A rotation by the angle $\theta$ along the axis $n_1\hat{x}+n_2\hat{z}$ can be realized by
 \begin{equation}
 e^{-\frac{i\theta}{2}(n_1\sigma_x+n_2\sigma_z)}=e^{-\frac{i\pi}{4}\sigma_x}  e^{-\frac{i\theta}{2}(n_1\sigma_x+n_2\sigma_y)}   e^{\frac{i\pi}{4}\sigma_x},
 \end{equation}
 while a similar rotation along the axis $n_1\hat{y}+n_2\hat{z}$ can be realized by
 \begin{equation}
 e^{-\frac{i\theta}{2}(n_1\sigma_x+n_2\sigma_z)}=e^{\frac{i\pi}{4}\sigma_y}  e^{-\frac{i\theta}{2}(n_1\sigma_y+n_2\sigma_x)}   e^{-\frac{i\pi}{4}\sigma_y}.
 \end{equation}
 \paragraph{}
 An alternative approach, which does not incorporate additional pulses over the theoretical sequence, involves applying the pulses off-resonantly. In this approach, driving the qubit transition (e.g. $m_s=0\leftrightarrow m_s=+1$) with energy $\omega_0$ at a Rabi frequency $\Omega$ is detuned by a frequency $\delta$. Under such a driving along the $x$ axis, $\Omega cos[(\omega_0+\delta)t]\sigma_x$, the total interaction Hamiltonian, in the rotation frame with respect to $\frac{1}{2}(\omega_0-\delta)\sum_a \sigma_z^{(a)}$ and under the rotating wave approximation yields
 \begin{equation}
 H=H_{NV}+H_{CTRL}=H_{NV}+\frac{\delta}{2}\sum_a \sigma_z^{(a)}+\frac{\Omega}{2}\sum_a \sigma_x^{(a)}.
 \end{equation}
 For a given Rabi frequency $\Omega$, a rotation along the axis $n_1\hat{x}+n_2\hat{z}$ can be realized by choosing
 \begin{equation}
 \delta=\frac{n_2}{n_1}\Omega.
 \end{equation} 
 Similarly, pulses along the axis $n_1\hat{y}+n_2\hat{z}$ can be introduced by driving with the same detuning along the $y$ axis (e.g. using in-phase-and-quadrature modulation).
 
\begin{figure}[!t]	
 	\includegraphics[width=1\columnwidth]{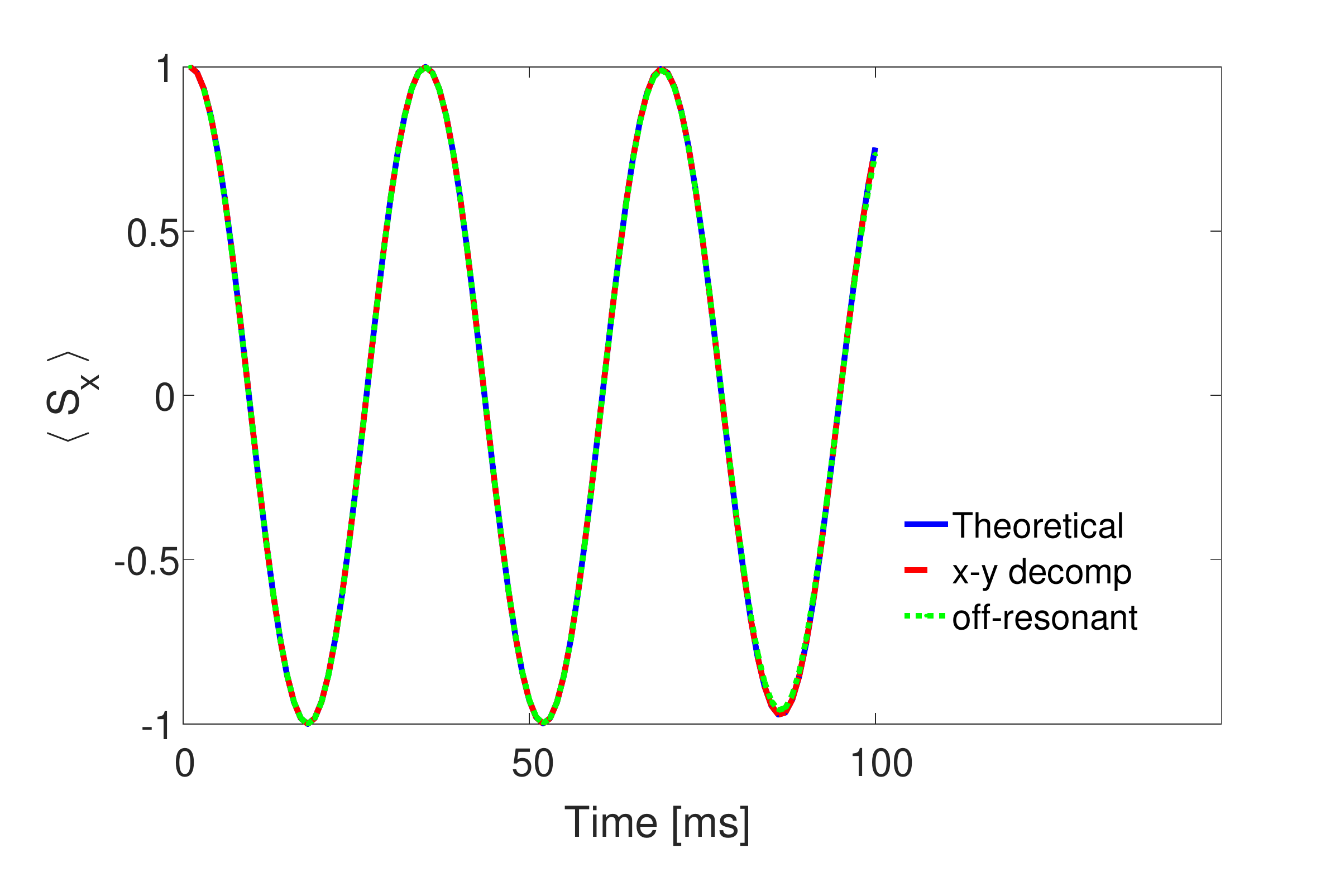}
 	\caption{(Color online) Simulated fidelity of an NV ensemble spin state initialized along the x axis as a function of time, under the evolution of the Hamiltonian generated by 100 repetitions of the icosahedral sequence presented in Fig. \ref{fig:sequences1} (c): ideal pulses, and experimental implementation approaches of 3-pulse-product decomposition an off-resonant pulses.}
 	\label{fig:expdynamics}
 \end{figure}

 \paragraph{}
 In order to estimate the experimental performance of the icosahedral sequences, we simulate the expected dynamics of an NV ensemble state with realistic parameters (464 spins with a typical interaction strength of $\sim 60$ Hz, see \cite{Farfurnik2018dip}) initialized along the $x$ axis under the icosahedral sequence in Fig. \ref{fig:sequences1} (c) using a cluster-based approach \cite{Farfurnik2018dip}. The similarities between the simulation results (presented in Fig. \ref{fig:expdynamics}) of applying theoretical instanteneous pulses and realistic pulses with $\Omega=10$ MHz, utilizing both experimental approaches described above, indicate the effectiveness of these sequences for Hamiltonian engineering. 
 \section{V. Conclusions}
 \paragraph{}
 To summarize, we use group theory considerations to generalize existing techniques for manipulating two-spin interaction terms in the context of average Hamiltonian theory. We then employ a linear programming code to extract the pulse sequence required for the engineering of target Hamiltonians from an initial ones. Such general manipulations can be fully implemented with pulses derived from icosahedral symmetry group, as opposed to Clifford rotations that provide partial capabilities [see Fig. \ref{fig:sequences2} (a)]. Starting from the NV-NV interaction Hamiltonian (general dipolar coupling), generatable target Hamiltonians include the Zeeman-preserving Hamiltonians essential for magnetic sensing (with enhanced sensitivity compared to Hamiltonians achievable using Clifford rotations) [Fig. \ref{fig:sequences1} (c)], Hamiltonians whose dynamics can be expressed analytically [Fig. \ref{fig:sequences2} (b)], as well as non-trivial Hamiltonians that could produce non-classical states (e.g. by modifying the WAHUHA sequence). Beyond enhanced magnetic sensing, this work could lead to novel studies of interacting spin ensembles. Substituting the constant external field $B_z$ in Eq. \eqref{eq:NVNV} by spin dependent coefficients $\{\delta^{(a)}_z\}$ may shed led on the effects of spin disorder and many-body localization \cite{Kucsko2018}, and form buliding blocks for solid-states quantum simulators \cite{Cappellaro2009,Farfurnik2018dip}.
\section*{Acknowledgements}
This work has been supported in part by the Minerva ARCHES award, the CIFAR-Azrieli global scholars program, the Israel Science Foundation (grant No. 750/14), the EU projects MetaboliQs (grant no. 820374) and PATHOS (grant no. 828946), the Ministry of Science and Technology, Israel, and the CAMBR fellowship for Nanoscience and Nanotechnology.

\appendix
\section{Appendix A: Hamiltonian Modification Utilizing Pulse Sequences - Theoretical Derivation}
\paragraph{}
A general three dimensional rotation by the angle $\theta$ around the axis $\hat{n}$ can be written in the form $R_{\hat{n},\theta}^{3\times 3}=\mathbb{I}+sin(\theta)[\hat{n}]_\times+[1-cos(\theta)][\hat{n}\otimes \hat{n}]$, where $\mathbb{I}$ is the $3 \times 3$ unity matrix and $[\hat{n}]_\times$ denotes the cross product of $\hat{n}$. For a specific rotation symmetry group with order $m$, a finite set of $m$ matrices $\{R_k^{3\times 3}\}$ represents all rotations applicable within the group. We denote by $\{R_k\}$ the generalized $2^N \times 2^N$ operators equally rotating all spins, and the matrix $R^{5\times 5}_k$ as the $5 \times 5$ reduced form of the matrix $R_k^{3\times 3} \otimes R_k^{3\times 3}$. The matrices in Eq. \eqref{eng} are given by 
\begin{align}
&\PS^{(1)}=\sum_k \frac{\tau_k}{T} R_k^{3\times 3},\\
&\PS^{(2)}=\sum_k \frac{\tau_k}{T} R^{5\times 5}_k,
\end{align} where $T$ is total cycle time and $\tau_k$ is the time duration within the sequence associated with the rotation $R_k$ in the following way: the k-th rotation element represents the product of all pulses applied before the time $t_k$, 
\begin{equation}
R_k=\prod_{i=1}^{k} P_i.
\end{equation}

\paragraph{}
For a pulse sequence $ \left\{  P_k, t_k \right\}$ (with spacings $\tau_j=t_j-t_{j-1}$), and following the notation \eqref{eng}, the superoperator \begin{equation}
\hat{\PS}[H]=\displaystyle\sum_{k=1}^n \frac{\tau_k}{T} R_k^{-1} H R_k
\end{equation}
characterizes the effect of the pulse sequence on the Hamiltonian via the Magnus expansion \cite{Magnus1954}. Applying this relation on the separate terms of the irreducible interaction form in Eqs. \eqref{general} and \eqref{eq:irreduc} yields
\begin{align*}
&\hat{\PS}(\vec{n}^{(a)}\cdot \vec{\sigma}^{(a)})=\left[\sum_k \frac{\tau_k}{T} \left(R_k^{3\times 3}\right)\vec{n}^{(a)}\right]\cdot \vec{\sigma}^{(a)}=\\
&\PS^{(1)}(\vec{n}^{(a)})\cdot \vec{\sigma}^{(a)}, \\\\
&\hat{\PS}(\alpha^{(ab)}\vec{\sigma}^{(a)}\cdot \vec{\sigma}^{(b)})=\alpha^{(ab)}\vec{\sigma}^{(a)}\cdot \vec{\sigma}^{(b)}, \\\\
&\hat{\PS}(\vec{\beta}^{(ab)}\cdot \vec{\lambda}_1^{(ab)})=\left[\sum_k \frac{\tau_k}{T} \left(R_k^{3\times 3}\right)\vec{\beta}^{(ab)}\right]\cdot \vec{\lambda}_1^{(ab)}=\\
&\PS^{(1)}(\vec{\beta}^{(ab)})\cdot \vec{\lambda}_1^{(ab)},\\\\
&\hat{\PS}(\vec{\gamma}^{(ab)}\cdot \vec{\lambda}_2^{(ab)})=\left[\sum_k \frac{\tau_k}{T} \left(R_k^{3\times 3}\otimes R_k^{3\times 3}\right)\vec{\gamma}^{(ab)}\right]\cdot \vec{\lambda}_2^{(ab)}=\\
&\PS^{(2)}(\vec{\gamma}^{(ab)})\cdot \vec{\lambda}_2^{(ab)}
\end{align*}

\bibliography{nvbibliography}
\end{document}